\documentclass[twocolumn,showpacs,preprintnumbers,amsmath,amssymb,aps,pra,groupedaddress,showpacs]{revtex4}
\usepackage{multirow,diagbox}  
\usepackage{siunitx}
\pdfoutput=1
\usepackage[colorlinks,citecolor=blue,linkcolor=red,hyperindex,CJKbookmarks]{hyperref}
\usepackage{amsmath}
\usepackage{graphicx}
\usepackage{dcolumn}
\usepackage{subfigure}
\usepackage{mathrsfs}
\usepackage{amssymb}
\usepackage{amsfonts}
\usepackage{makecell}
\usepackage[dvipsnames]{xcolor}

\begin{document}
\draft
\title{Criticality-Enhanced Quantum Sensing in the Anisotropic Quantum Rabi Model}
\author{Xin Zhu, Jia-Hao L\"{u}, Wen Ning, Fan Wu}
\author{Li-Tuo Shen}\thanks{E-mail: lituoshen@yeah.net}
\author{Zhen-Biao Yang}\thanks{E-mail: zbyang@fzu.edu.cn}
\author{Shi-Biao Zheng}
\affiliation{Fujian Key Laboratory of Quantum Information and Quantum Optics,\\
	College of Physics and Information Engineering, Fuzhou University, Fuzhou,\\
	Fujian 350108, China}
\date{\today }
\date{\today }
\begin{abstract}
Quantum systems that undergo quantum phase transitions exhibit divergent susceptibility and can be exploited as probes to estimate physical parameters. We generalize the dynamic framework for criticality-enhanced quantum sensing by the quantum Rabi model (QRM) to its anisotropic counterpart and derive the correspondingly analytical expressions for the quantum Fisher information (QFI). We find that the contributions of the rotating-wave and counterrotating-wave interaction terms are symmetric at the limit of the infinite ratio of qubit frequency to field frequency, with the QFI reaching a maximum for the isotropic quantum Rabi model. At finite frequency scaling, we analytically derive the inverted variance of higher-order correction and find that it is more affected by the rotating-wave coupling than by the counterrotating-wave coupling.

\end{abstract}

\date{\today}

\maketitle
\section{INTRODUCTION}
The importance of precision measurement in physics and other sciences has made it a long-pursued goal for the vast majority of scientific researchers. Compared to classical precision measurements, quantum metrology can significantly enhance the sensitivity of parameter estimation by utilizing the distinct features of quantum effects such as entanglement~\cite{1}, squeezing~\cite{2} and quantum criticality~\cite{3,4,5,6,7}. For instance, the properties of the equilibrium state may change significantly when the physical parameters change slightly near the critical point. This sensitivity of critical behavior provides a powerful resource for estimating physical parameters.

In the past few years, quantum criticality has attracted growing interest and many quantum sensing protocols based on quantum criticality have been proposed~\cite{8,9,10,11,12,13,14,15,16,17,18,19,20,21,22}. However, the time required to prepare the ground state close to the critical point also diverges, which obliterates the advantages provided by quantum criticality. Recent results suggest that the stringent requirement for state preparation can be relaxed by the dynamical method~\cite{23}, showing that a divergent scaling in the quantum Fisher information (QFI) can be achieved for general initial states by taking the QRM as an explicit example. The work of Ref.~\cite{23} dealt with the standard (isotropic) QRM, in which the rotating and counterrotating terms have the uniform coupling strengths. However, in real light-matter interaction systems, the available coupling strengths are typically much smaller than the frequencies, so that the counter-rotating--wave terms have a negligible effect on the system dynamics. A promising strategy to overcome this restriction is to introduce a parametric drive, which can effectively transform the rotating--wave coupling into an asymmetric combination of rotating-- and counter-rotating--wave couplings~\cite{add1,add2}. As such, an investigation of criticality-enhanced sensing based on the anisotropic quantum Rabi Model (AQRM) is of practical relevance.

Recent years have witnessed impressive explorations on different aspects of the AQRM, such as the enhanced squeezing~\cite{24}, the QFI~\cite{25}, and the quantum phase transition (QPT)~\cite{26,27,28,29}. To date, several creative schemes have been proposed to realize the AQRM, including the ones utilizing a two-dimensional quantum well~\cite{25}, trapped ions~\cite{30}, and superconducting circuit~\cite{31,32,33}. 

In this article, we investigate quantum sensing in the AQRM and analyze the scope of application under the situations of different anisotropic ratios between rotating-wave and counterrotating-wave interaction terms. We analytically derive the formula of the QFI for the AQRM, and its divergent scaling. We find that the rotating-wave and counterrotating-wave interaction terms have symmetrical effects on the critically-enhanced quantum sensing at the limit of the infinite ratio of qubit frequency to field frequency. However, at finite frequency scaling, the effect of higher order correction will break this equilibrium and shift the inverted variance maxima located in the QRM in the direction where the rotating-wave interaction prevails.

\section{The QFI of critical quantum dynamics}
The performance of quantum sensing is related to the sensitivity of state with respect to the change of a parameter that can be quantified by QFI, which is introduced by extending the classical Fisher information to quantum regime~\cite{34,35}. The QFI for the estimation of the parameter $\alpha$ has a relatively simple form as
\begin{eqnarray}\label{Eq1}
	\mathscr{F}_\alpha= 4{\rm{Var}}\left[h_\alpha\right]_{\left\vert\Psi\right\rangle},
\end{eqnarray}
where $\rm{Var}\left[\ldots\right]_{\left\vert\Psi\right\rangle}$ is variance related to the initial state  $\left\vert\Psi\right\rangle$ and $h_\alpha =
iU^{\dagger}_{\alpha}\partial_{\alpha}U_{\alpha} =-i\partial_{\alpha}U^{\dagger}_{\alpha}U_{\alpha}$ is the transformed local generator of parametric translation of $U_{\alpha} = exp\left(-iH_{\alpha}T\right)$ with respect to $\alpha$\textcolor{red}{~\cite{36+,37,38}}, where $H_{\alpha} = H_{0}+\alpha H_{1}$ represents a family of  parameter $(\alpha)$-dependent Hamiltonians, whose eigenvalue equation satisfies the following form~\cite{37}
\begin{eqnarray}\label{Eq2}
	\left[H_{\alpha},\Gamma\right] = \sqrt{\Delta}\Gamma,
\end{eqnarray}
where $\Gamma=i\sqrt{\Delta}A-B$ with $A=-i\left[H_0,H_1\right]$, $B=-\left[H_{\alpha},\left[H_0,H_1\right]\right]$ and $\Delta$ is dependent on the parameter $\alpha$. This type of Hamiltonian $H_{\alpha}$ may have an isometric energy spectrum in which the energy gap $\varepsilon\thicksim\sqrt{\Delta}$ when $\Delta>0$, and $\sqrt{\Delta}$ becomes imaginary if $\Delta<0$. The normal-to-superradiant phase transition occurs at the critical point defined by $\Delta=0$. 
The transformed local generator can be written as 
\begin{eqnarray}
	h_{\alpha} = H_1t + \frac{{\rm{cos}}(\sqrt{\Delta}t)-1}{\Delta}A - \frac{{\rm{sin}}(\sqrt{\Delta}t)-\sqrt{\Delta}t}{\Delta\sqrt{\Delta}t}B.
\end{eqnarray}
It can be seen that $h_{\alpha}$ becomes divergent as $\Delta\to0$ if $\sqrt{\Delta}t\backsimeq\mathcal{O}(1)$, which represents a signature of critical quantum dynamics. Substituting the transformed local generator into Eq.~(\ref{Eq1}), we obtain the QFI as follows:

\begin{eqnarray}\label{Eq4}
	\mathscr{F}_{\alpha}(t)\backsimeq4\frac{[{\rm{sin}}(\sqrt{\Delta}t)-\sqrt{\Delta}t]^{2}}{\Delta^3}{\rm{Var}}\left[B\right]_{\left\vert\Psi\right\rangle}.
\end{eqnarray}
It obviously shows that the QFI diverges under the condition of $\sqrt{\Delta}t\backsimeq\mathcal{O}(1)$. The requirement for ground state preparation can be avoided based on the fact that this scaling of the QFI results from the dynamic  evolution of quantum system itself and is applicable to general initial state $ \left\vert\Psi\right\rangle $ provided that $ {\rm{Var}}\left[B\right]_{\left\vert\Psi\right\rangle}\backsimeq\mathcal{O}(1) $ or general mixed state. The prominent feature for the cases proposed here is that any Hamiltonian satisfying Eq.~(\ref{Eq1}) takes effect when it's applied to such a kind of quantum sensing~\cite{23}.

\section{Critical quantum sensing in the AQRM}
The AQRM describes the interaction between a qubit and a bosonic field mode with asymmetric rotating- and counterrotating-wave coupling strengths. The system dynamics is governed by the Hamiltonian $(\hbar =1)$
\begin{eqnarray}
	\mathcal{H} &=&\omega a^{\dagger}a + \frac{\Omega}{2}\sigma_{z}+\lambda_{1}(a\sigma_{+} +a^{\dagger}\sigma_{-}) \nonumber\\
	&&+ \lambda_{2}(a\sigma_{-} +a^{\dagger}\sigma_{+}),
\end{eqnarray}
where $\sigma_{\pm}=\left(\sigma_{x} \pm \emph{i}\sigma_{y}\right)/2$ and $\sigma_{z}$ are Pauli operators of the qubit with the transition frequency $\Omega$, and $a^{\dagger}$ $(a)$ is a creation (annihilation) operator for the bosonic field with the frequency $\omega$. $\lambda_{1}$ and $\lambda_{2}$ respectively characterize the coupling strengths of rotating-wave and counterrotating-wave interactions between the qubit and the bosonic field. The ratio $\lambda_1/\lambda_2$ depicts distinctive feature of the AQRM, as compared to the isotropic one that has been investigated in the context of criticality-enhanced quantum sensing~\cite{23}. Without loss of
generality, we set $ \lambda_{1} $ and $ \lambda_{2} $ to be real. 

In the limit of $\eta =\Omega/\omega \rightarrow \infty$, we can obtain an effective low-energy Hamiltonian,

\begin{eqnarray}\label{Eq6}
	\mathcal{H}_{np}^{\downarrow} &=& \left(\omega - \frac{\lambda_{1}^2+\lambda_{2}^2}{\Omega}\right)a^{\dagger}a-\frac{\lambda_{1}\lambda_{2}}{\Omega}(a^{\dagger2}+a^2) \nonumber\\
	&&-\frac{\lambda_{2}^2}{\Omega}-\frac{\Omega}{2}.
\end{eqnarray}
Eq.~(\ref{Eq6}) can be diagonalized as $H_{np}^{\downarrow} = \varepsilon_{np}a^{\dagger}a + E_{np}$, with the energy gap $\varepsilon_{np} = \omega\sqrt{\Delta_{g}}/2$, and the ground-state energy $E_{np}=\frac{1}{2}\left(\varepsilon_{np}-\omega+\frac{\lambda_{1}^{2}-\lambda_{2}^{2}}{\Omega}-\Omega\right)$, where $\Delta_g=4(1-g^2)(1-\gamma^2g^2)$, $g = (\lambda_{1}+\lambda_{2})/\sqrt{\omega\Omega}$, and $\gamma = (\lambda_{1}-\lambda_{2})/(\lambda_{1}+\lambda_{2})$. The energy gap $\varepsilon_{np}$ is real only when $g<1$ and vanishes at $g=1$, locating the critical point at which QPT happens. 
For this Hamiltonian, the QFI regarding estimation of the parameter g is
\begin{small}
\begin{eqnarray}\label{Eq7}
	\mathscr{F}_g(t)\backsimeq16g^{2}\xi^{2}\mu^{2}\frac{[{\rm{sin}}(\sqrt{\Delta_{g}}\omega t)-\sqrt{\Delta_{g}}\omega t]^{2}}{\Delta_{g}^3}{\rm{Var}}\left[P^2\right]_{\left\vert\varphi\right\rangle_c},
\end{eqnarray}	
\end{small}where $\xi=\gamma^{2}-1$, $\mu=1-\gamma^2g^2$, $P = i(a^{\dagger} - a)/\sqrt{2}$ and $\left\vert\varphi\right\rangle_c$ is the initial state of the bosonic field (see Appendix \ref{A} for a proof). It can be found that when $g\to1$ (i.e., $\Delta_{g}\to0$), $\mathscr{F}_{g}\to\infty$. This would allow us to estimate the parameter with a precision enhanced by critical quantum dynamics. The quantum Cram\'{e}r-Rao bound (QCRB)~\cite{39} associated with the QFI characterizes how well a parameter can be estimated from a probability distribution and gives the ultimate precision of the quantum parameter estimation. Here, we study quantum sensing based on such an AQRM and explore two experimentally feasible measurement methods to achieve the precision of the same order as QCRB. 

\section{Measurement schemes for AQRM-based sensing}
The first method is based on quadrature measurements of the bosonic field by standard homodyne detection~\cite{40}. Without losing generality, the system is assumed to be initially in a product state of qubit's $\left\vert\downarrow\right\rangle_{q}$ state and field's photon superposition state $\left\vert\varphi\right\rangle_c$: $\left\vert\Psi\right\rangle = \left\vert\downarrow\right\rangle_{q}\otimes\left\vert\varphi\right\rangle_c$, with $\left\vert\varphi\right\rangle_c = (\left\vert0\right\rangle+i\left\vert1\right\rangle)/\sqrt{2}$. After an evolution for time $t$ under the Hamiltonian, the mean value of the quadrature X, defined as $X = (a + a^{\dagger})/\sqrt{2}$, is
\begin{eqnarray}
		\left\langle{X}\right\rangle_{t} = \sqrt{2}\Delta_{g}^{-\frac{1}{2}}\mu {\rm{sin}}(\sqrt{\Delta_{g}}\omega t/2),
\end{eqnarray}
with the variance
\begin{eqnarray}
	\left(\Delta X\right)^{2} = 1 -2g^{2}\xi^{2}\mu^{2}\Delta_{g}^{-1}\left[1-{\rm{cos}}\left(\sqrt{\Delta_{g}}\omega t\right)\right].
\end{eqnarray} 
\begin{figure}[t]
	\centering
	\includegraphics[width=0.5\textwidth]{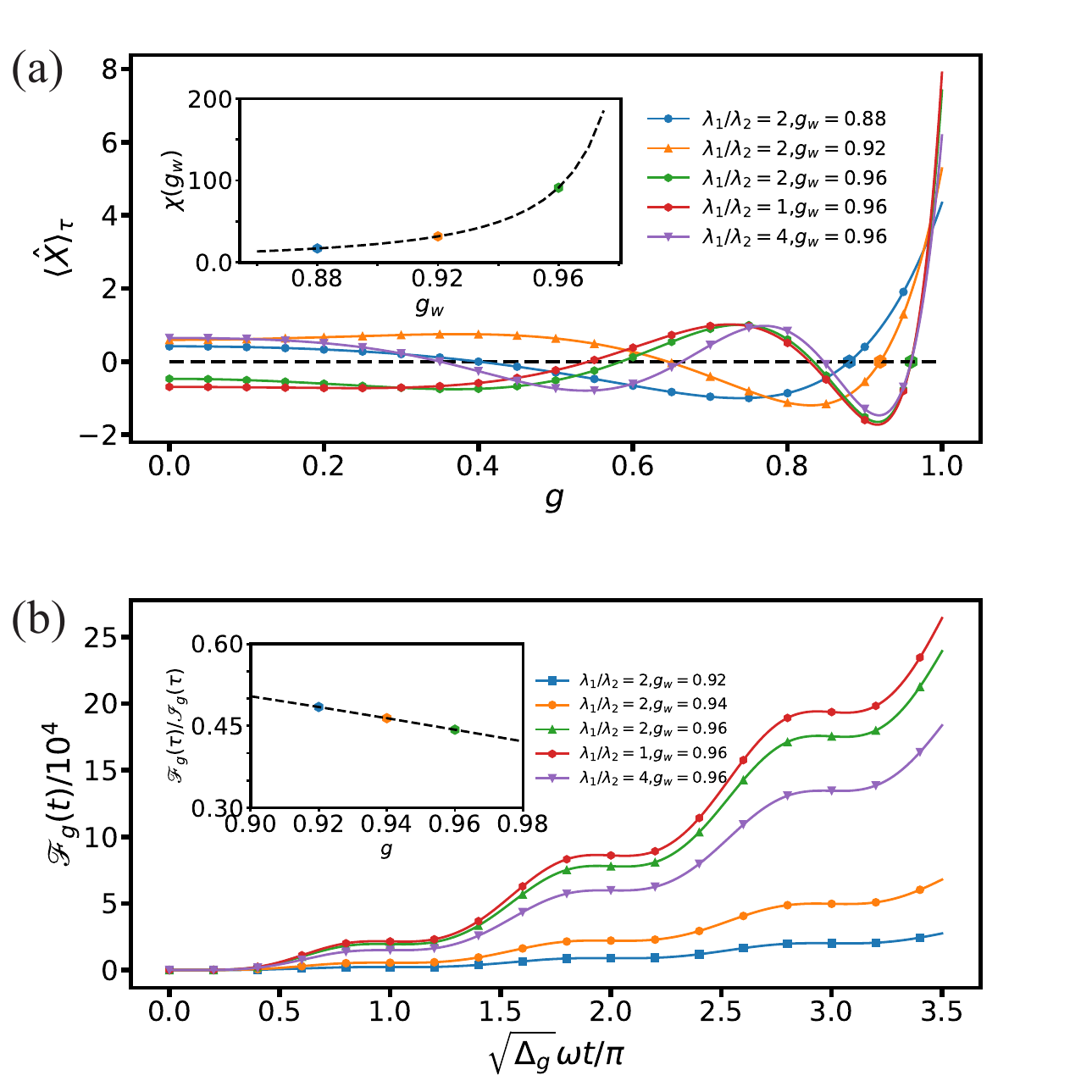}
	\caption{(a) Mean value of quadratures $X$ at different working points during an evolution time $\tau=2\pi/(\sqrt{\Delta_{g_w}}\omega)$ as a function of $g$. The inset shows the corresponding susceptibility at different working points $\chi(g_w)\equiv\chi_{g}(\tau)\vert_{g=g_w}=4\sqrt{2}\pi g_w\Delta_{g_w}^{-3/2}$ with $\lambda_{1}/\lambda_{2}=2$. (b) QFI $\mathscr{F}_g(t)$ as a function of the evolution time $t$. Inset: the local maximum of the inverted variance $\mathscr{I}_g(\tau)$ after an evolution time $\tau=2\pi/(\sqrt{\Delta_{g}}\omega)$ reaches the same order of $\mathscr{F}_g(\tau)$ with $\lambda_{1}/\lambda_{2}=2$.
	}
	\label{f1}
\end{figure}The inverted variance defined as
\begin{eqnarray}
	\mathscr{I}_{g}=\chi_{g}^{2}(t)/\left(\Delta X\right)^{2},
\end{eqnarray}
can be used to quantify the precision of the parameter estimation, where $\chi_{g}(t) = \partial_{g}\left\langle{X}\right\rangle_{t}$ is the susceptibility of $\left\langle{X}\right\rangle_{t}$ with respect to the parameter $g$, and exhibits a divergent behavior when $g\to1$ for different anisotropic ratio $\lambda_1/\lambda_2$ (Fig.~\ref{f1}(a)), an analogous feature appearing in the standard QRM with $\lambda_1/\lambda_2=1$. The precision of the parameter estimation reaches the QCRB when $\mathscr{I}_{g}(t)=\mathscr{F}_{g}(t)$. The inverted variance achieves its local maximum as
\begin{eqnarray}
	\mathscr{I}_{g}(\tau_{k})\backsimeq32\pi^{2}g^{2}\mu^{4}\Delta_{g}^{-3}k^{2}
\end{eqnarray}
(see Appendix \ref{B} for a proof), at $t = \tau_{k} = 2k\pi/\left(\sqrt{\Delta_{g}}\omega\right)\left(k\in\mathbb{Z}^{+}\right)$. The QFI at the same time can be obtained from Eq.~(\ref{Eq7}) as
\begin{eqnarray}
		\mathscr{F}_{g}(\tau_{k})\backsimeq 64\pi^{2}g^{2}\xi^{2}\mu^{2}\Delta_{g}^{-3}k^{2}{\rm{Var}}\left[P^2\right]_{\left\vert\varphi\right\rangle_c}.
\end{eqnarray}
It can be seen from Fig.~\ref{f1}(b) that $\mathscr{I}_{g}(\tau_{k})$ can reach the same order of $\mathscr{F}_g(\tau_k)$ for different anisotropic ratio $\lambda_1/\lambda_2$, 
though it is reduced with the increase of $\lambda_1/\lambda_2$. Note that the optimal case happens for the isotropic condition $\lambda_1/\lambda_2 = 1$.
We stress that the results of the AQRM considered here, like that of the QRM, also holds without requiring particular initial states of the bosonic field~\cite{23}.
\begin{figure}[t]
	\centering
	\includegraphics[width=0.5\textwidth]{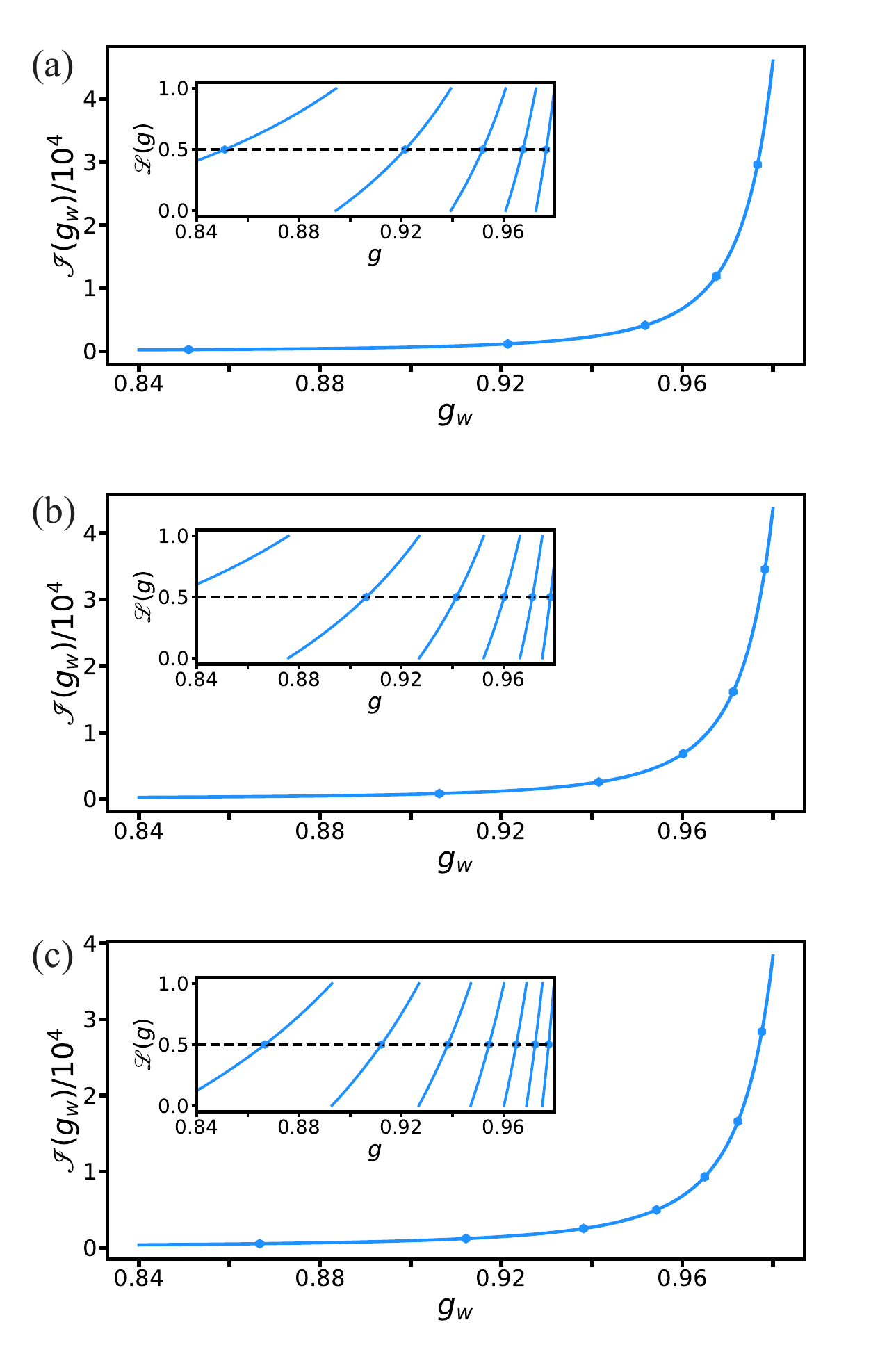}     
	\caption{Inverted variance $\mathscr{I}(g_w)$ after evolution time $\tau=4\pi/(\sqrt{\Delta_{g_w}}\omega)$. The inset shows the selected working points under the condition $\mathcal{L}(g)=0.5$. The initial state is $\left\vert\Psi\right\rangle =(c_{\uparrow}\left\vert\uparrow\right\rangle_{q} +c_{\downarrow}\left\vert\downarrow\right\rangle_{q})\otimes\left\vert\varphi\right\rangle_c$. Without loss of generality, we assume $2c_{\uparrow}^{*}c_{\downarrow}=1$ and $\left\vert\varphi\right\rangle_c =\left\vert0\right\rangle $. As three examples, we consider the three cases: (a) $\lambda_{1}/\lambda_{2}=1$, (b) $\lambda_{1}/\lambda_{2}=2$ and (c) $\lambda_{1}/\lambda_{2}=4$.}
	\label{f2}
\end{figure}

Another method is to directly measure the qubit to extract the information on the parameter. Without loss of generality~\cite{41,42,43}, we assume that the initial state of the system is a product state  $\left\vert\Psi\right\rangle =(c_{\uparrow}\left\vert\uparrow\right\rangle_{q} +c_{\downarrow}\left\vert\downarrow\right\rangle_{q})\otimes\left\vert\varphi\right\rangle_c$. The mean value of the qubit's observable $\langle\sigma_{x}\rangle$ is
\begin{eqnarray}
	\langle\sigma_{x}\rangle =2{\rm{Re}}[c_{\uparrow}^{*}c_{\downarrow}\mathcal{G}(g,t)],
\end{eqnarray}
where $\mathcal{G}(g,t) = \langle\varphi\vert u_{\uparrow}^{\dagger}u_{\downarrow}\vert\varphi\rangle_{c}$ is the Loschmidt amplitude~\cite{42}, and $u_{\uparrow}$($u_{\downarrow}$) is the evolution operator of bosonic field when the qubit in state $\left\vert\uparrow\right\rangle_{q}$ ($\left\vert\downarrow\right\rangle_{q}$). The inverted variance corresponding to $\langle\sigma_{x}\rangle$ is 
\begin{eqnarray}
	\mathscr{I}_{g}=\left(\partial_{g}\langle\sigma_{x}\rangle\right)^{2}/{\rm{Var}}[\sigma_{x}].
\end{eqnarray}
In Fig.~\ref{f2}, we assume $2c_{\uparrow}^{*}c_{\downarrow}=1$ and plot the inverted variance $\mathscr{I}_{g_w}=\mathscr{I}(\tau)\vert_{g=g_w}$ at the working point with an evolution time $\tau=4\pi/(\sqrt{\Delta_{g_w}}\omega)$ for  the estimation of the parameter $g$ based on the observable $\langle\sigma_{x}\rangle$ with three anisotropic ratios. The point that satisfies $\mathcal{L}(g_w)=0.5$ is selected as the working point, where $\mathcal{L}=L(g)-\lfloor L(g)\rfloor$ with $L(g) =\left\{[(1+g^{2})(1+\gamma^{2}g^{2})]/[(1-g^{2})(1-\gamma^{2}g^{2})]\right\}^{1/2}$. Under this condition, the mean value of the observable is $\langle\sigma_{x}\rangle\backsimeq0$.
It can be found that $\mathscr{I}_{g_w}$ scales as $\Delta_{g_w}^{-3}$, which shows a divergent feature close to the critical point. It is worth noting that we can obtain similar results from other general initial states, such as coherent states and the superposition of Fock states (see Appendix \ref{C}). Obviously, as compared to the homodyne detection with the QRM (Fig.~\ref{f2}(a)), for the cases of the AQRM and with the increase of the anisotropic ratio $\lambda_1/\lambda_2$, there are more working points satisfying $\mathcal{L}(g_w)=0.5$, as shown in Fig.~\ref{f2}(b), (c).


To quantify the performance of the present sensing protocol, we use the Ramsey interferometry as the benchmark, which works by sandwitching a free evolution with a time $\tau$ between two Ramsey pulses, each performing a $\pi/2$ rotation on the qubit, whose frequency is to be estimated. The qubit, starting with the initial state $\left\vert\downarrow\right\rangle$, has a probability of being populated in $\left\vert\uparrow\right\rangle$, given by
\begin{eqnarray}
	P_{\uparrow} = (1/2)(1+ {\rm{cos}}\theta),
\end{eqnarray}
where $\theta = \omega_q \tau$, with $\omega_q$ characterizes the qubit's transition frequency. Such a probability is related to the Bloch vector  $\sigma_{z}$ by $P_{\uparrow}= (1+\left\langle\sigma_{z}\right\rangle)/2$. The interferometer is most susceptible to the variation of $\omega_q$ around the bias point~\cite{qs}, where $\theta_o=(n+1/2)\pi$ and the susceptibility is
\begin{eqnarray}
	\left\vert\partial P_{\uparrow}/\partial\theta\right\vert_{\theta=\theta_o}=1/2.
\end{eqnarray}
This implies that a longer evolution time is preferred for improving the susceptibility, which, however, would introduce more serious decoherence noises. At this point, the standard deviation associated with measurement of $P_{\uparrow}$ is $\Delta P_{\uparrow}=1/2$, which leads to inverted variance
\begin{eqnarray}
	(\partial P_{\uparrow}/\partial\theta)^2/(\Delta P_{\uparrow})^2=1.
\end{eqnarray}
In distinct contrast with this result, inverted variance in the present protocol exhibits a much higher inverted variance near the critical point, as shown in Fig.~\ref{f2}. We note that both sensing protocols work within a limited parameter range. For the Ramsey interferometry, the detection has a limited linear range with $\left\vert\delta\omega_q\tau\right\vert<\pi/2$, where $\delta\omega_q$ is the deviation of qubit frequency from the reference point and $\tau$ denotes the free evolution time~\cite{qs}. When exceeding such a range, phase wrapping occurs, breaking down the one-to-one correspondence between the transition probability and the qubit frequency.

\section{Finite-frequency scaling} 
We now turn to investigating the finite-frequency effect. We derive a high-order correction to the effective low-energy Hamiltonian up to fourth order in $\omega/\Omega$: 
\begin{eqnarray}
	\mathcal{H}_{np}^{\Omega}&=&\mathcal{H}_{np}^{\downarrow} + \frac{1}{\Omega^{3}}(DC)^2- \frac{\omega}{\Omega^2}(\lambda_{1}^2a^{\dagger}a - \lambda_{2}^2aa^{\dagger}),
\end{eqnarray}
where $C=\lambda_{1}a + \lambda_{2}a^{\dagger}$, $D =\lambda_{2}a + \lambda_{1}a^{\dagger}$, and the high-order correction makes the Hamiltonian no longer contain only even terms for the bosonic field compared with Eq.~(\ref{Eq6}). We find that the odd order terms will break the equilibrium between the rotating-wave and counterrotating-wave interactions and shift the maximum point of the inverted variance at the QRM towards the direction where the rotating-wave coupling prevails. The inverted variance of the high-order correction 
by homodyne detection of the bosonic field mode can be written as
\begin{eqnarray}\label{Eq16}
	\mathscr{I}_{lab}^g(\tau) &=& \frac{g^2\pi^2}{2(1-g^2)^3} + \frac{g^2\pi^2(2+g^4)}{(1-g^2)^4\eta}\gamma \nonumber\\
	&&+ \frac{g^2\pi^2E}{4(1-g^2)^6\eta^2}\gamma^2 + \mathcal{O}(\gamma^3),
\end{eqnarray}
where $E = 4g^{10} + 4(2+\eta^2) + 2g^{8}(3\eta^2-8) - 2g^2(9\eta^2+4) - 2g^6(11\eta^2 + 10)+g^4(30\eta^2+32-\pi ^2)$ (see Appendix \ref{D}). It can be seen from Fig.~\ref{f_bias} that the shift of the maximum point of the inverted variance will increase when $g\to1$ and as the limit $\Omega/\omega \to \infty$  is not satisfied, the influence of $\eta$ is non-negligible.

\begin{figure}[]
	\centering
	\includegraphics[width=0.5\textwidth]{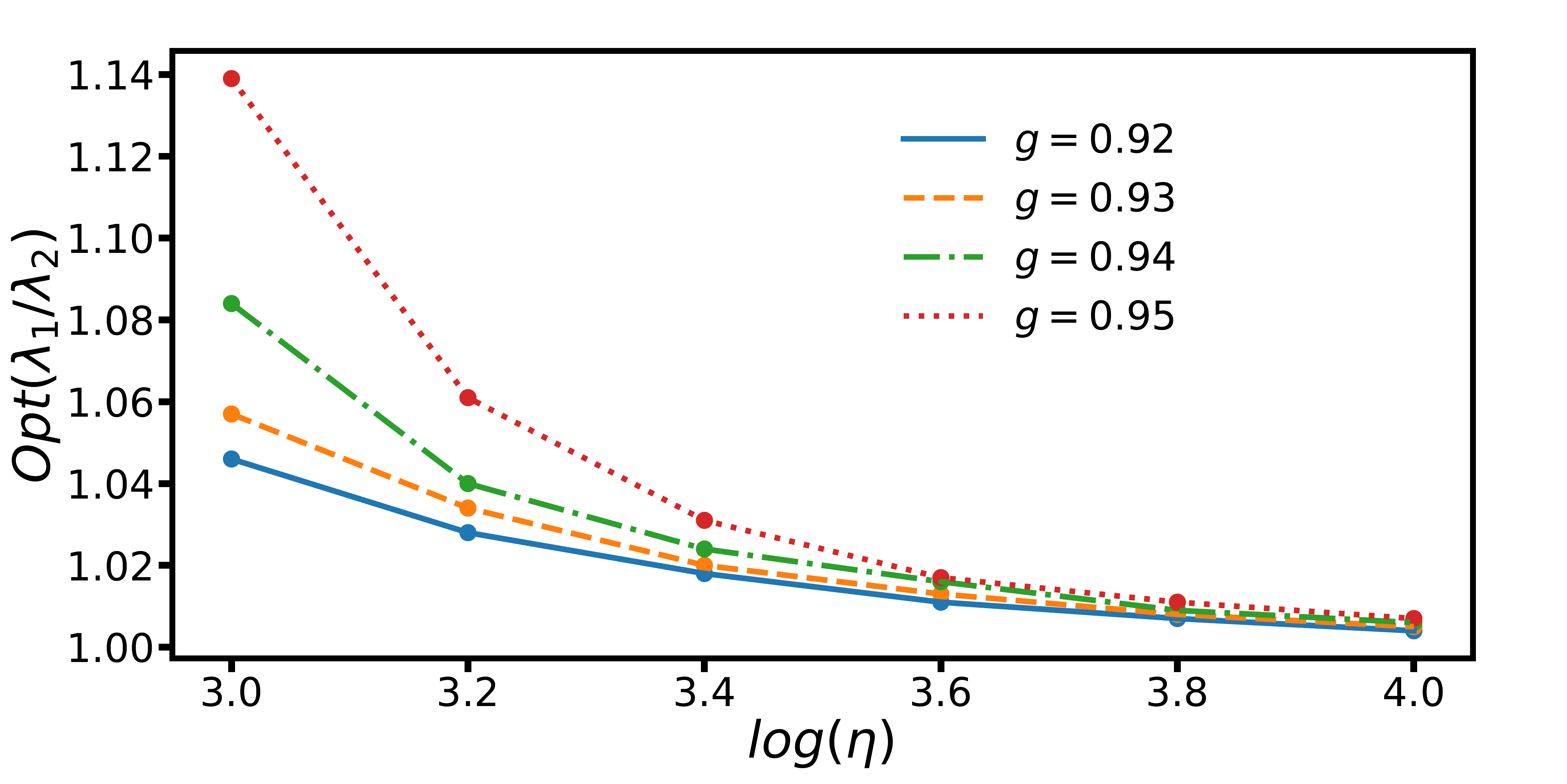}
	\caption{Optimal ratio of $\lambda_{1}/\lambda_{2}$ at the maximum point of the QFI under finite frequency ratio. The evolution time is not exactly $\tau=2\pi/\left(\sqrt{\Delta_{g}}\omega\right)$ and $\Delta_g=4(1-g^2)(1-\gamma^2g^2)$ is not applicable for finite frequency. We choose the maximum of $\mathscr{I}_{lab}^g(t)$ as the result for which evolution time $t$ is $0.9\tau\le t\le1.0\tau$.}
	\label{f_bias}
\end{figure}

\section{Conclusion}
To summarize, we have investigated the criticality-enhanced quantum sensing based on the dynamical evolution of the AQRM. We show the divergent scaling of the QFI under different anisotropic ratios, which weakens as the anisotropy ratio increases. We find that the influences of  the rotating-wave and counterrotating-wave interaction terms are symmetric when the ratio of the qubit transition frequency to the field frequency is infinite. For a finite frequency ratio, the equilibrium between the rotating-wave and counterrotating-wave interaction is broken, and a bias in the couplings favors improvement of the precision. 

\section*{ACKNOWLEDGMENTS}
This work was supported by the National Natural Science Foundation of China (Grant No. 12274080, No. 11874114, No. 11875108), the National Youth Science Foundation of China (Grant No. 12204105), the Educational Research Project for Young and Middle-aged Teachers of Fujian Province (Grant No. JAT210041), and the Natural Science Foundation of Fujian Province (Grant No. 2021J01574, No. 2022J05116).
\appendix\section{The QFI of the AQRM for general initial states}\label{A}
We map the AQRM in the normal phase into a block-diagonal form by making Schrieffer-Wolff transformation,
\begin{eqnarray}
	H^{’} &\backsimeq& e^{-\mathcal{S}}\mathcal{H}e^{\mathcal{S}} \nonumber\\
	&=& \frac{1}{\Omega}\left[CD-\frac{1}{\Omega^2}(CD)^2-\frac{\omega}{\Omega}(\lambda_{2}^2a^{\dagger}a - \lambda_{1}^2aa^{\dagger})\right]\sigma_{+}\sigma_{-} \nonumber\\
	&&- \frac{1}{\Omega}\left[DC-\frac{1}{\Omega^2}(DC)^2+\frac{\omega}{\Omega}(\lambda_{1}^2a^{\dagger}a - \lambda_{2}^2aa^{\dagger})\right]\sigma_{-}\sigma_{+} \nonumber\\
	&&+\omega a^{\dagger}a + \frac{\Omega}{2}\sigma_{z},
\end{eqnarray}
in which 
\begin{eqnarray}
	\mathcal{S} &\backsimeq&\frac{1}{\Omega}\left(C\sigma_{+}-D\sigma_{-}\right)-\frac{4}{3\Omega^3}\left(CDC\sigma_{+}-DCD\sigma_{-}\right) \nonumber\\
	&& +\frac{\omega}{\Omega^2}\left[\left(\lambda_{1}a-\lambda_{2}a^{\dagger}\right)\sigma_{+} +\left(\lambda_{2}a-\lambda_{1}a^{\dagger}\right)\sigma_{-}\right].
\end{eqnarray}

Projecting the $H^{’}$ (as $\eta \to \infty$) to the spin-down subspace, we obtain
\begin{eqnarray}
		H_{np}^{\downarrow}
		&=& \frac{\omega}{2}\left[(X^2 + P^2)-g^2(X^2+\gamma^2P^2)\right],
\end{eqnarray}
with the quadrature operators defined as $X = (a + a^{\dagger})/\sqrt{2}$ and $P = i(a^{\dagger} - a)/\sqrt{2}$. We choose $H_0 = \omega(X^2 + P^2)/2$ and $H_1 = \omega(X^2 + \gamma^2P^2)/2$, the corresponding A and B can be written as 
\begin{eqnarray}
	A &=& \frac{\omega^2}{2}(\gamma^2-1)(XP+PX),\\
	B &=& \omega^3(\gamma^2-1)\left[(1-g^2)X^2-(1-\gamma^2g^2)P^2\right].
\end{eqnarray}
By defining $\alpha = -g^2$ and applying the method in Ref. ~\cite{23} with $\Delta=4\omega^2(1-g^2)(1-\gamma^2g^2)$, we can obtain the QFI for the measurement of the parameter $g$ as Eq.~(\ref{Eq7}).

\section{ Calculation details of quadrature dynamics of the AQRM}\label{B}
In the Heisenberg picture, the mean value of the quadrature $X$ at time $t$ can be written as follows

	\begin{eqnarray}
		\left<X\right>_{t} &=& \left\langle\Psi\right\vert\exp(iH_{np}^{\downarrow}/t)X\exp(-iH_{np}^{\downarrow}/t)\left\vert\Psi\right\rangle\nonumber\\  &=&\sum_{n=0}^{\infty}\frac{(i\omega t/2)^n}{n!}\left[H_{np},X\right]_n \nonumber\\ 
		&=&\sqrt{2}\Delta_{g}^{-\frac{1}{2}}\mu {\rm{sin}}(\sqrt{\Delta_{g}\omega t/2})P\nonumber\\ 
		&&+{\rm{cos}}(\sqrt{\Delta_{g}\omega t/2})X,
	\end{eqnarray}

where $\left\vert\Psi\right\rangle$ is the  initial state, $\left[H_{np}^{\downarrow},X\right]_{n+1}=\left[H_{np}^{\downarrow},\left[H_{np}^{\downarrow},X\right]_n\right]$ and $\left[H_{np}^{\downarrow},X\right]_0 = X$. In the main text, we set the system to be initialized in $\left\vert\varphi\right\rangle_c = (\left\vert0\right\rangle+i\left\vert1\right\rangle)/\sqrt{2}$. The dynamics
of the quadrature $X$ can be obtained as 
\begin{eqnarray}
	\left<X\right>_{t} = \sqrt{2}\frac{\mu}{\sqrt{\Delta_{g}}}{\rm{sin}}(\sqrt{\Delta_{g}\omega t/2}),
\end{eqnarray}
from which we obtain the susceptibility with respect to the field frequency $g$ as

	\begin{eqnarray}\label{Eq17}
		\chi_{g}(t) &=& \partial_{g}\left\langle{X}\right\rangle_{t} \nonumber\\
		&=& -\frac{1}{\sqrt{2}}(4\mu^2\Delta_{g}^{-1}+\gamma^2)g\omega t{\rm{cos}}(\sqrt{\Delta_{g}\omega t/2})\nonumber\\
		&&+4\sqrt{2}\mu g\xi \Delta_{g}^{-3/2}{\rm{sin}}(\sqrt{\Delta_{g}\omega t/2}).
	\end{eqnarray}

\begin{figure}[t]
	\centering
	\includegraphics[width=0.5\textwidth]{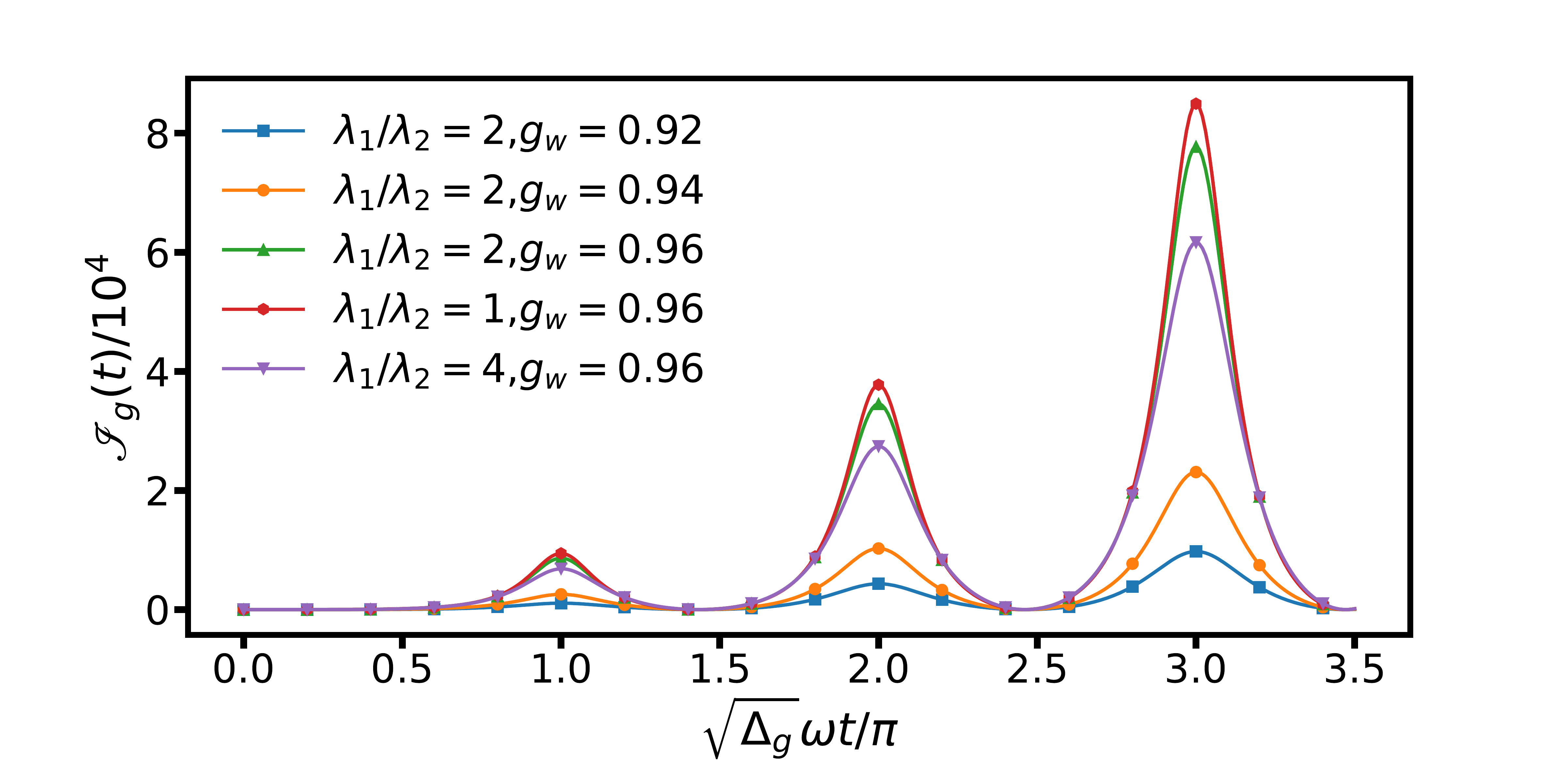}
	\caption{The inverted variance $\mathscr{I}_g(t)$ as a function of evolution time $t$. The inverted variance $\mathscr{I}_g(t)$ reaches a local maximum $\mathscr{I}_g(t)\backsimeq8g^{2}\omega^{2}\mu^{4}\Delta_{g}^{-2}t^{2}$ periodically at $\sqrt{\Delta_{g}}\omega t/(2\pi) =\sqrt{(1-g^{2})(1-\gamma^{2}g^{2})}\omega t/\pi \in\mathbb{Z}^{+} $. The peak of the inverse variance has a certain width, which allows for small deviations of the evolution time while the inverse variance still maintains a large value.}
	\label{f4}
\end{figure}

The divergent behavior occurs as $\Delta_{g}\to0$. In particular, after an evolution time $t = \tau_{k} = 2k\pi/\left(\sqrt{\Delta_{g}}\omega\right)\left(k\in\mathbb{Z}^{+}\right)$ the first term of $\chi_{g}$ is zero, so we have
\begin{eqnarray}
	\chi_{g} = (-1)^{n-1}\frac{1}{\sqrt{2}}(4\mu^2\Delta_{g}^{-1}+\gamma^2)g\omega\tau_{k}.
\end{eqnarray}
As the same way, we get the mean value of the quadrature $X^2$ at time $t$ as

\begin{eqnarray}
	\left<X^2\right>_{t} = 1-2\mu\xi g^2\Delta_{g}^{-1}[1-{\rm{cos}}(\sqrt{\Delta_{g}\omega t})].
\end{eqnarray}
Hence, the variance of the quadrature $X$ is 
\begin{small}
	\begin{eqnarray}
		(\Delta X)^2 &=& \left<X^2\right>_{t} - \left<X\right>_{t}^{2} \nonumber\\
		&=& 1-\mu(2g^2\xi+\mu)\Delta_{g}^{-1}[1-{\rm{cos}}(\sqrt{\Delta_{g}\omega t})].
	\end{eqnarray}
\end{small}It can be seen that the oscillation factor $1-{\rm{cos}}(\sqrt{\Delta_{g}\omega t})$ in the second term of $(\Delta X)^2$ is out phase of ${\rm{cos}}(\sqrt{\Delta_{g}\omega t/2})$ in $\chi_{g}(t)$ from Eq.~(\ref{Eq17}). This means that we can enhance susceptibility while limiting the  fluctuation of the quadrature in a small range, which allows the measurement precision of the parameter $g$ to be significantly improved. In Fig.~\ref{f4}, we plot the curve of the  inverted variance $\mathscr{I}_{g}=\chi_{g}^{2}(t)/\left(\Delta X\right)^{2}$ with different parameters, which reaches its local maximums $\mathscr{I}_g(t)\backsimeq8g^{2}\omega^{2}\mu^{4}\Delta_{g}^{-2}t^{2}$ periodically at $\sqrt{\Delta_{g}}\omega t/(2\pi) =\sqrt{(1-g^{2})(1-\gamma^{2}g^{2})}\omega t/\pi \in\mathbb{Z}^{+} $. We note here that, our results through the AQRM follow the same rules as the QRM does.
\section{Proof of measurement}\label{C}
In the main text, we choose $2c_{\uparrow}^{*}c_{\downarrow}=1$ as an example and the mean value of the local observable of the qubit can be simplified as $\langle\sigma_{x}\rangle ={\rm{Re}}[\mathcal{G}(g,t)]$. The inverted variance corresponding to $\langle\sigma_{x}\rangle$ is given by
\begin{eqnarray}
	\mathscr{I}_{g}=\frac{(\partial_{g}\langle\sigma_{x}\rangle)^{2}}{1-\langle\sigma_{x}\rangle^2}=\frac{{\rm{Re}}[\chi^{\mathcal{G}}(g,t)]^2}{1-{\rm{Re}}[\mathcal{G}(g,t)]^2}.
\end{eqnarray}
In order to choose the appropriate working point, we project the initial state $\vert\varphi\rangle_{c}$ of the bosonic field 
into the eigenspaces of $H_{np}^{\downarrow} = \left(\omega - \frac{\lambda_{1}^{2}+\lambda_{2}^{2}}{\Omega}\right)a^{\dagger}a -\frac{\lambda_{1}\lambda_{2}}{\Omega}\left(a^{\dagger2}+a^{2}\right)$ and $H_{np}^{\uparrow} = \left(\omega + \frac{\lambda_{1}^{2}+\lambda_{2}^{2}}{\Omega}\right)a^{\dagger}a +\frac{\lambda_{1}\lambda_{2}}{\Omega}\left(a^{\dagger2}+a^{2}\right) $ respectively,
\begin{eqnarray}
	\vert\varphi\rangle_{c} = \sum_{n}c_{m}^{\downarrow}\vert\psi_{m}^{\downarrow}\rangle,\vert\varphi\rangle_{c} = \sum_{n}c_{m}^{\uparrow}\vert\psi_{m}^{\uparrow}\rangle,
\end{eqnarray}
where $\vert\psi_{m}^{\sigma}\rangle = S(r_{\sigma})\vert m\rangle,S(r_{\sigma}) = \exp[(r_{\sigma}/2)(a^{\dagger2}-a^{2})]$ with $r_{\downarrow} = -\frac{1}{4}
\ln\left[1-\frac{4\lambda_{1}\lambda_{2}}{\omega\Omega-(\lambda_{1}-\lambda_{2})^2}\right]$ and $r_{\uparrow} = -\frac{1}{4}
\ln\left[1+\frac{4\lambda_{1}\lambda_{2}}{\omega\Omega+(\lambda_{1}-\lambda_{2})^2}\right]$. Hence, the Loschmidt amplitude can be rewritten as
	\begin{eqnarray}
		\mathcal{G}(g,t) &=& \langle\varphi\vert u_{\uparrow}^{\dagger}u_{\downarrow}\vert\varphi\rangle_{c}\nonumber\\ &=&\sum_{m,n}c_{m}^{\uparrow*}c_{n}^{\downarrow}\exp[i\omega t(m\sqrt{(1+g^2)(1+\gamma^2g^2)}\nonumber\\
		&&-n\sqrt{(1-g^2)(1-\gamma^2g^2)})]\langle\psi_{m}^{\uparrow}\vert\psi_{m}^{\downarrow}\rangle.
	\end{eqnarray}
In particular, when $\tau = 2\pi/[\omega\sqrt{(1-g^2)(1-\gamma^2g^2)}]$
, we find that
	\begin{eqnarray}
		e^{-iH_{np}^{\downarrow}\tau}\vert\varphi\rangle_{c} &=&  \sum_{n}c_{n}^{\downarrow}e^{-i2\pi n}\vert\psi_{n}^{\downarrow}\rangle 
		= \vert\varphi\rangle_{c},
	\end{eqnarray}
which leads to a solution of 
	\begin{eqnarray}
		\mathcal{G}(g,\tau)&=&\langle\varphi\vert e^{iH_{np}^{\uparrow}\tau}\vert\varphi\rangle_{c}\nonumber\\
		&=&\sum_{m,n}c_{m}^{\uparrow*}c_{n}^{\downarrow}e^{i\omega \tau m\sqrt{(1+g^2)(1+\gamma^2g^2)}}\langle\psi_{m}^{\uparrow}\vert\psi_{m}^{\downarrow}\rangle \nonumber\\
		&=& \sum_{m}\vert c_{m}^{\uparrow}\vert^2e^{2im\pi L(g)} \nonumber\\
		&=&\sum_{m}\vert c_{m}^{\uparrow}\vert^2e^{2im\pi\mathcal{L}(g)}.
	\end{eqnarray}
It can be found $\mathcal{G}(g,\tau)=\sum_{m}(-1)^{m}\vert c_{m}^{\uparrow}\vert^{2}$ under the condition $\mathcal{L}(g)=0.5$, which is approximately zero if the coefficient $c_{m}^{\uparrow}$ varies very slowly as $m$ increases. Thus, we choose the working points $g_w$ which satisfies $\mathcal{L}(g_w)=0.5$  and the associated evolution time as $\tau = 4\pi/(\omega\Delta_{g_w}^{1/2})=2\pi/[\omega\sqrt{(1-g_w^2)(1-\gamma^2g_w^2)}]$. 
The corresponding inverted variance can be approximated as
\begin{eqnarray}
	\mathscr{I}(g_w)&\backsimeq& {\rm{Re}}[\chi^{\mathcal{G}}(g_w,\tau)]^2\nonumber\\
	&\backsimeq& 64\pi^2g_w^2\xi^2\mu^2\Delta_{g_w}^{-3}\upsilon^2,
\end{eqnarray}
where $\upsilon = Im\left[\langle\varphi\vert u_{\uparrow}^{\dagger}u_{\downarrow}P^2\vert\varphi\rangle_{c}\right]$.
\begin{figure}[]
	\centering
	\includegraphics[width=0.5\textwidth]{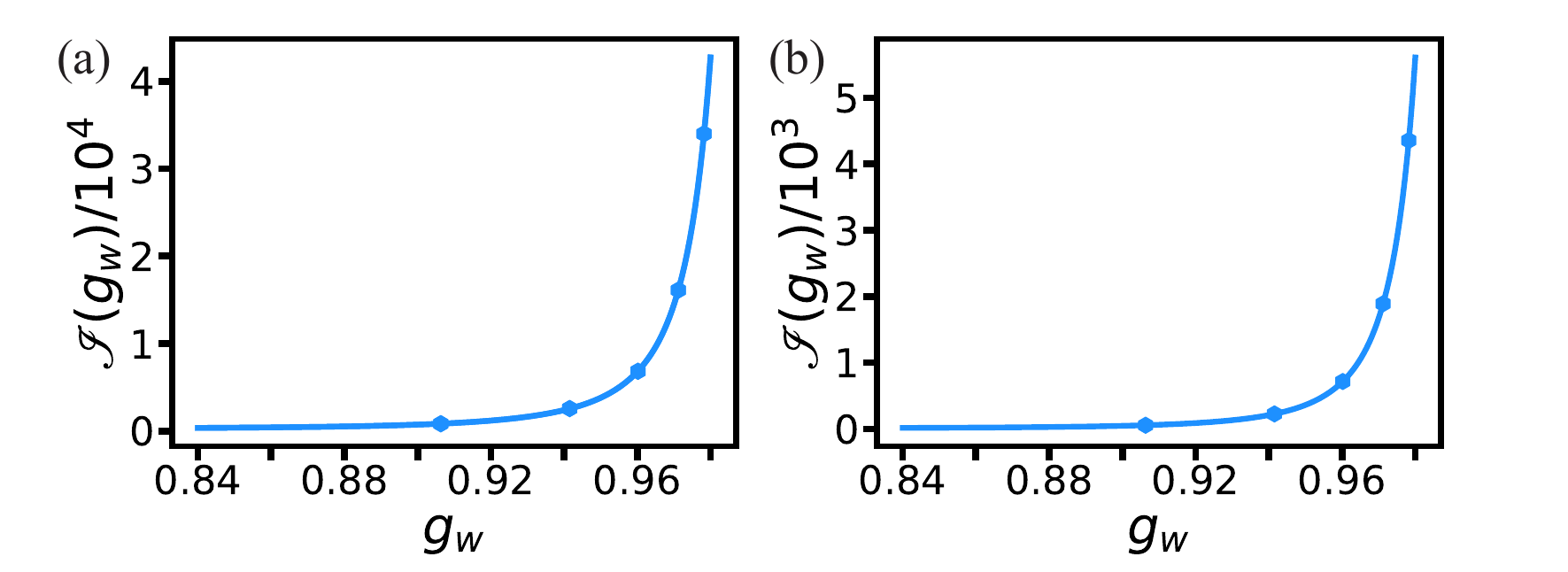}    
	\caption{The working points are the same as Fig.~\ref{f2}(a). In those two cases, (a) superposition of Fock states $\left\vert\varphi\right\rangle_c = (\left\vert0\right\rangle+\left\vert1\right\rangle)/\sqrt{2}$, (b) coherent state $\left\vert\alpha\right\rangle$ both exhibit divergent behavior. }
	\label{f5}
\end{figure}

To further prove that this divergent behavior holds for general initial states, we choose a superposition of Fock states $\left\vert\varphi\right\rangle_c = (\left\vert0\right\rangle+\left\vert1\right\rangle)/\sqrt{2}$ and a coherent state $\left\vert\alpha\right\rangle$ with $\alpha=1$ as two examples, see Fig.~\ref{f5}.

\section{Finite-frequency scaling}\label{D}
Our previous analysis is under the limit of $\eta = \Omega/\omega\to\infty$, which seems to play the analogous role of a thermodynamic limit. However, this remains a challenge in experiment. We investigate the influence of the finite-frequency which is described by the following low-energy effective Hamiltonian as
\begin{eqnarray}
	\mathcal{H}_{np}^{\Omega}&=& \langle\downarrow\vert H^{’}\vert\downarrow\rangle    \nonumber\\
	&=& \mathcal{H}_{np}^{\downarrow}+ \frac{1}{\Omega^{3}}(DC)^2- \frac{\omega}{\Omega^2}(\lambda_{1}^2a^{\dagger}a - \lambda_{2}^2aa^{\dagger})\nonumber\\
	&=&-\frac{g^2\omega}{4}\eta^{-1}(X^2 + \gamma^2P^2 - \gamma)^2 \nonumber\\ 
	&& -\frac{g^2\omega}{2}\eta^{-1}\gamma(X^2 +P^2) + H_{np}^{\downarrow},
\end{eqnarray}
in which the first two terms add the odd terms of $\gamma$. The present of odd terms will disrupt the balance of the rotating-wave and counterrotating-wave interaction, which also leads to a shift in the point of maximum value located in the inverse variance of the QRM. After retaining  the quadratic potential of $\gamma$, we obtain the corresponding inverted variance as (i.e.~Eq. (\ref{Eq16}) in the main text)
\begin{eqnarray}
	\mathscr{I}_{lab}^g(\tau) &=&\chi_{g}^{2}(t)/\left(\Delta X\right)^{2} \nonumber\\
	&=& \frac{g^2\pi^2}{2(1-g^2)^3} + \frac{g^2\pi^2(2+g^4)}{(1-g^2)^4\eta}\gamma \nonumber\\
&&+ \frac{g^2\pi^2E}{4(1-g^2)^6\eta^2}\gamma^2 + \mathcal{O}(\gamma^3).
\end{eqnarray}


\begin{thebibliography}{99}
\bibitem{1}R. Horodecki, P. Horodecki, M. Horodecki, and K. Horodecki, Quantum entanglement, Rev. Mod. Phys. 81, 865 (2009).
\bibitem{2} C. M. Caves, Quantum-mechanical noise in an interferometer. Phys. Rev. D 23, 1693 (1981).
\bibitem{3}S. Sachdev, Quantum Phase Transitions (Cambridge University Press, Cambridge, England, 2011).
\bibitem{4}K. Macieszczak, M. Guţă, I. Lesanovsky, and J. P. Garrahan,
Dynamical phase transitions as a resource for quantum
enhanced metrology, Phys. Rev. A 93, 022103 (2016).
\bibitem{5}M. M. Rams, P. Sierant, O. Dutta, P. Horodecki, and J.
Zakrzewski, At the Limits of Criticality-Based Quantum
Metrology: Apparent Super-Heisenberg Scaling Revisited,
Phys. Rev. X 8, 021022 (2018).
\bibitem{6}I. Fr\'{e}rot and T. Roscilde, Quantum Critical Metrology,
Phys. Rev. Lett. 121, 020402 (2018).
\bibitem{7}P. Zanardi, M. G. A. Paris, and L. Campos Venuti, Quantum
criticality as a resource for quantum estimation, Phys. Rev.
A 78, 042105 (2008).
\bibitem{8} G. N. Gol’tsman, O. Okunev, G. Chulkova, A. Lipatov, A.
Semenov, K. Smirnov, B. Voronov, A. Dzardanov, C.
Williams, and R. Sobolewski, Picosecond superconducting
single-photon optical detector, Appl. Phys. Lett. 79, 705
(2001).
\bibitem{9}M. Raghunandan, J. Wrachtrup, and H. Weimer, HighDensity Quantum Sensing with Dissipative First Order
Transitions, Phys. Rev. Lett. 120, 150501 (2018).
\bibitem{10} N. Wang, G.-Q. Liu, W.-H. Leong, H. Zeng, X. Feng, S.-H.
Li, F. Dolde, H. Fedder, J. Wrachtrup, X.-D. Cui, S. Yang,
Q. Li, and R.-B. Liu, Magnetic Criticality Enhanced Hybrid
Nanodiamond Thermometer Under Ambient Conditions,
Phys. Rev. X 8, 011042 (2018).
\bibitem{11}L.-P. Yang and Z. Jacob, Quantum critical detector:
Amplifying weak signals using discontinuous quantum
phase transitions, Opt. Express 27, 10482 (2019).
\bibitem{12} S. Dutta and N. R. Cooper, Critical Response of a
Quantum van der Pol Oscillator, Phys. Rev. Lett. 123,
250401 (2019).
\bibitem{13}P. A. Ivanov, Steady-state force sensing with single trapped
ion, Phys. Scr. 95, 025103 (2020).
\bibitem{14} L. Garbe, M. Bina, A. Keller, M. G. A. Paris, and S.
Felicetti, Critical Quantum Metrology with a Finite Component Quantum Phase Transition, Phys. Rev. Lett.
124, 120504 (2020).
\bibitem{15}M. Tsang, Quantum transition-edge detectors, Phys. Rev. A 88, 021801(R) (2013).
\bibitem{16}K. Macieszczak, M. Guţă, I. Lesanovsky, and J. P. Garrahan, Dynamical phase transitions as a resource for quantum enhanced metrology, 
Phys. Rev. A 93, 022103 (2016).
\bibitem{17}P. Zanardi, M. G. A. Paris, and L. C. Venuti, Quantum criticality as a resource for quantum estimation, Phys. Rev. A
78, 042105 (2008).
\bibitem{18}M. Bina, I. Amelio, and M. G. A. Paris, Dicke coupling by feasible local measurements at the superradiant quantum phase transition, Phys. Rev. E 93, 
052118 (2016).
\bibitem{19}S. Fernández-Lorenzo and D. Porras, Quantum sensing close to a dissipative phase transition: Symmetry breaking and criticality as metrological resources, Phys. Rev. A 96,
013817 (2017).
\bibitem{20}M. M. Rams, P. Sierant, O. Dutta, P. Horodecki, and J.
Zakrzewski, At the Limits of Criticality-Based Quantum Metrology: Apparent Super-Heisenberg Scaling Revisited, Phys. Rev. X 8, 021022 (2018).
\bibitem{21}T. L. Heugel, M. Biondi, O. Zilberberg, and R. Chitra, Quantum Transducer Using a Parametric Driven-Dissipative Phase Transition, Phys.
Rev. Lett. 123, 173601 (2019).
\bibitem{22}P. A. Ivanov and D. Porras, Adiabatic quantum metrology with strongly correlated quantum optical systems, Phys. Rev. A 88, 023803 (2013).
\bibitem{23}Y. Chu, S. Zhang, B. Yu, and J. Cai, Dynamic Framework for Criticality-Enhanced Quantum Sensing, Phys. Rev. Lett. 126,
010502 (2021).
\bibitem{add1}W. Qin, A. Miranowicz, P. B. Li, X. Y. Lü, J. Q. You, and
F. Nori, Exponentially Enhanced Light-Matter Interaction, Cooperativities, and steady-state Entanglement Using Parametric
Amplification, Phys. Rev. Lett. 120, 093601 (2018).
\bibitem{add2}C. Leroux, L. C. G. Govia, and A. A. Clerk, Enhancing Cavity
Quantum Electrodynamics via Antisqueezing: Synthetic Ultrastrong Coupling, Phys. Rev. Lett. 120, 093602 (2018).
\bibitem{24}Y.-Y. Zhang and X.-Y. Chen, Analytical solutions by squeezing to the anisotropic Rabi model in the nonperturbative deep-strong-coupling regime, Phys. Rev. A 96, 063821 (2017).
\bibitem{25}Z.-H. Wang, Q. Zheng, X. Wang and Y. Li, The energy-level crossing behavior
and quantum Fisher information in a quantum well with spin-orbit coupling, Sci. Rep. 6 22347 (2016).
\bibitem{26}M.-J. Hwang, R. Puebla, and M. B. Plenio, Quantum Phase
Transition and Universal Dynamics in the Rabi Model,
Phys. Rev. Lett. 115, 180404 (2015).
\bibitem{27}M. Liu, S. Chesi, Z. J. Ying, X. Chen, H. G. Luo, and
H. Q. Lin, Universal Scaling and Critical Exponents of the
Anisotropic Quantum Rabi Model, Phys. Rev. Lett. 119,
220601 (2017).
\bibitem{28}M. J. Hwang and M. B. Plenio, Quantum Phase Transition in the Finite Jaynes-Cummings Lattice Systems, Phys. Rev. Lett. 117,
123602 (2016).
\bibitem{29}L. T. Shen, Z. B. Yang, H. Z. Wu, and S. B. Zheng, Quantum
phase transition and quench dynamics in the anisotropic Rabi
model, Phys. Rev. A 95, 013819 (2017).
\bibitem{30} I. Aedo and L. Lamata, Analog quantum simulation of generalized Dicke models in trapped ions. Phys. Rev. A 97, 042317 (2018).
\bibitem{31}G. Wang, R. X, H.-Z. Shen, C. Sun and K. Xue, simulating Anisotropic quantum
Rabi model via frequency modulation, Sci. Rep.9 4569 (2019).
\bibitem{32}Y. Wang, W.-L. You, M. Liu, Y.-L. Dong, H.-G. Luo, G. Romero and J.-Q. You, Quantum criticality and state engineering in the simulated anisotropic quantum Rabi model, New J. Phys. 20 053061 (2018).
\bibitem{33}Q.-T. Xie, S. Cui, J.-P. Cao, L. Amico, and H. Fan, Anisotropic Rabi Model, Phys. Rev. X 4, 021046 (2014).
\bibitem{34}Fisher, R. A. Theory of statistical estimation. Math. Proc. Cambridge Philos. Soc. 22, 700–725 (1925).
\bibitem{35}S. Pang, A. N. Jordan, Optimal adaptive control for quantum metrology with time-dependent Hamiltonians, Nature Communications volume 8, 14695 (2017).
\bibitem{36}Braunstein, S. L., Caves, C. M. and Milburn, G. J. Generalized uncertainty relations: theory, examples, and Lorentz invariance. Ann. Phys. 247, 135–173 (1996).
\bibitem{36+}Giovannetti, V., Lloyd, S. and Maccone, L. Quantum metrology. Phys. Rev. Lett.
96, 010401 (2006).

\bibitem{37}S. Pang and T. A. Brun, Quantum metrology for a
 general Hamiltonian parameter, Phys. Rev. A 90, 022117
 (2014).
\bibitem{38}S. S. Pang and A. N. Jordan, Optimal adaptive control for
quantum metrology with time-dependent Hamiltonians,
Nat. Commun. 8, 14695 (2017).
\bibitem{39}C. W. Helstrom, Quantum Detection and Estimation Theory (Academic Press, New York, 1976); See also S. D. Personick, IEEE Trans. Inf. Theory 17, 240 (1971); H. P. Yuen and M. Lax, ibid. 19, 740 (1973); A. S. Holevo, Probabilistic and Statistical Aspects of Quantum Theory (North-Holland, Amsterdam, 1982).
\bibitem{40}K. Banaszek and K. Wódkiewicz, Operational theory of homodyne detection, Phys. Rev. A 55, 3117 (1997).
\bibitem{41} H. T. Quan, Z. Song, X. F. Liu, P. Zanardi, and C. P. Sun,
Decay of Loschmidt Echo Enhanced by Quantum Criticality, Phys. Rev. Lett. 96, 140604 (2006).
\bibitem{42}Z. P. Karkuszewski, C. Jarzynski, and W. H. Zurek,
Quantum Chaotic Environments, the Butterfly Effect,
and Decoherence, Phys. Rev. Lett. 89, 170405 (2002).
\bibitem{43}Y. Liu, J. Tian, R. Betzholz, and J. Cai, Pulsed QuantumState Reconstruction of Dark Systems, Phys. Rev. Lett. 122, 110406 (2019).
\bibitem{qs}C. L. Degen, F. Reinhard, and P. Cappellaro, Quantum sensing,
Rev. Mod. Phys. 89, 035002 (2017).

\end{thebibliography}
\end{document}